\documentclass[prl,showpacs,twocolumn,amsmath,amssymb,superscriptaddress]{revtex4-1}

\usepackage{graphicx}
\usepackage{dcolumn}
\usepackage{bm}
\usepackage{times}
\usepackage{color}

\newcommand{\anferm}{\psi^{\phantom{\dagger}}}
\newcommand{\crferm}{\psi^{\dagger}}

\newcommand{\momx}{\hat{\pi}_x}
\newcommand{\momy}{\hat{\pi}_y}

\begin{document}


\title{
Switchable Quantum Anomalous Hall state in a strongly frustrated lattice magnet}

\author{J\"orn W.F. Venderbos}
\affiliation{IFW Dresden, P.O. Box 27 01 16, D-01171 Dresden, Germany}

\author{Maria Daghofer}
\affiliation{IFW Dresden, P.O. Box 27 01 16, D-01171 Dresden, Germany}

\author{Jeroen {van den Brink}}
\affiliation{IFW Dresden, P.O. Box 27 01 16, D-01171 Dresden, Germany}

\author{Sanjeev Kumar}
\affiliation{Indian Institute of Science Education and Research (IISER) Mohali, Knowledge city, Sector 81, Mohali 140306, India
}

%
\begin{abstract}
We establish that the interplay of itinerant fermions with localized magnetic
moments on a checkerboard lattice leads to magnetic flux-phases. 
For weak itineracy the flux-phase is coplanar and the electronic dispersion takes the
shape of graphene-like Dirac fermions. 
Stronger itineracy drives the formation of a non-coplanar, chiral flux-phase, in which the Dirac fermions acquire a topological mass
that is proportional to a ferromagnetic spin polarization. 
Consequently the system self-organizes into a ferromagnetic Quantum Anomalous Hall state in which the direction of its dissipationless edge-currents can be switched by an applied magnetic field.
\end{abstract}


\maketitle

{\it Introduction.---} 
The study of topologically non-trivial states of matter is one of the
hottest topics in present day condensed matter physics. An
understanding of topological states requires a theoretical paradigm
that goes far beyond the concept of global symmetry breaking that has
originally been laid out by Landau. It is remarkable that the
theoretical predictions on the existence of various topologically
ordered states have rather swiftly led to the discovery of an entirely
new class of materials, the topological
insulators~\cite{Kane:2005gb,Roy:2009kk,Moore:2007gq,Fu:2007io}. Recent
pioneering experiments have confirmed the key signatures of
non-trivial topology in certain materials, e.g. spin-momentum-locked undoubled
 Dirac fermions~\cite{Xia:2008tc,Hsieh:2009dp,Xia:2009ii} and the
Quantum Spin Hall (QSH) effect~\cite{Koenig:2007hs}.  
These topological insulators are time-reversal (TR) invariant
generalizations of the first, much older, topological state of matter,
the famous Integer Quantum Hall states~\cite{VONKLITZING:1980wf,TKNN} 
that are induced by a magnetic field, which obviously breaks TR
symmetry. 

In a seminal work in 1988, Haldane established that a magnetic field
is not required to induce states with the same topology as IQH states~\cite{HALDANE:1988uf}.
It was shown that adding complex 
hopping to a graphene-like Hamiltonian for electrons on a
honeycomb lattice opens up topologically nontrivial gaps at the Dirac points,
which yields a topologically ordered, insulating state, 
referred to as a Quantum Anomalous Hall (QAH) state.
An  important feature of QAH states are edge channels,
in which current can run only in one direction; in contrast to QSH
states, on a single edge the opposite spin channel carrying the opposite 
current is absent~\cite{Liu:2008ev}. QAH states would thus allow very robust,
dissipationless charge transport along edge channels, as
backscattering would be completely suppressed. However, while
  signatures of QAH behavior have been reported in some
  compounds~\cite{Taguchi:2001fx,machida_QAH_07,AHE_PdCrO2_2010}, the 
  QAH state is the only one among these topologically
insulating states that remains to be unambiguously
identified in experiment.

The experimental difficulty is mirrored by the 
frailty of  theoretical mass-generating mechanisms for a graphene-like kinetic
energy with a linear dispersion at the Fermi level. 
TR-symmetry breaking via (magnetic) order requires rather
specific and strong longer-range Coulomb
interactions~\cite{Raghu:2008kr}, because the Dirac cones' vanishing density
of states at the Fermi level renders interaction-driven
ordered states energetically less favorable. QAH states can  more readily be
induced in models with a finite density of
states~\cite{Martin:2008dx,Kumar:2010de,Ohgushi:2000tu}, especially in
cases of quadratic band crossings~\cite{Sun:2009je}, as for instance found in the checkerboard
lattice, which exhibit a weak-coupling instability~\cite{Sun:2009je,Sun:2011dk,Uebelacker:2011bb}.
Another
approach has been to consider spin-orbit coupled magnetic semi-conductors \cite{Qi:2006jm} or spin-polarized QSH states~\cite{Liu:2008ev}.

Our starting point is instead the Kondo
lattice model, which provides the most general context for the study of the interplay between 
localized spins and itinerant electrons. 
We will show that, depending on parameters, its ground state on the checkerboard lattice can feature
massless Dirac cones or a chiral QAH state. Moreover, because the spin texture
underlying the QAH state has a net ferromagnetic (FM) moment, we have a direct
switching mechanism between ground-states with different chirality just by flipping the FM polarization. 
The ensuing possibility to reverse the direction of an edge current by an external magnetic field is an attractive feature in the context of spintronics. 

In particular we focus on the case of electrons strongly coupled to large localized spins
that in turn interact via a strongly frustrated antiferromagnetic
superexchange. 
In the absence of charge carriers, the
magnetic interactions give rise to a highly degenerate ground state
manifold (GSM) comprised of all spin configurations that obey certain
\emph{local} constraints, which go by the name of {\it spin-ice
  rules}, as a reference to three-dimensional spin systems that remain
disordered down to the lowest temperatures like the H protons in water
ice~\cite{Bramwell:2001ex}. It is clear that in a frustrated
spin-systems governed by ice rules, doping of itinerant charge
carriers will (partly) lift the macroscopic the degeneracy of the
GSM~\cite{Jaubert:2012vi} because of the kinetic energy competing with
the local ice-rule constraints. We will show that for not too large
kinetic energies, a unique flux-phase ground-state is selected in our
case, whose low-energy states are described by a massless Dirac
equation. When the kinetic energy becomes stronger, enhancing the
competition, these Dirac fermions acquire a topological mass, and the
ferromagnetic QAH state emerges.

\begin{figure}
\includegraphics[width=\columnwidth]{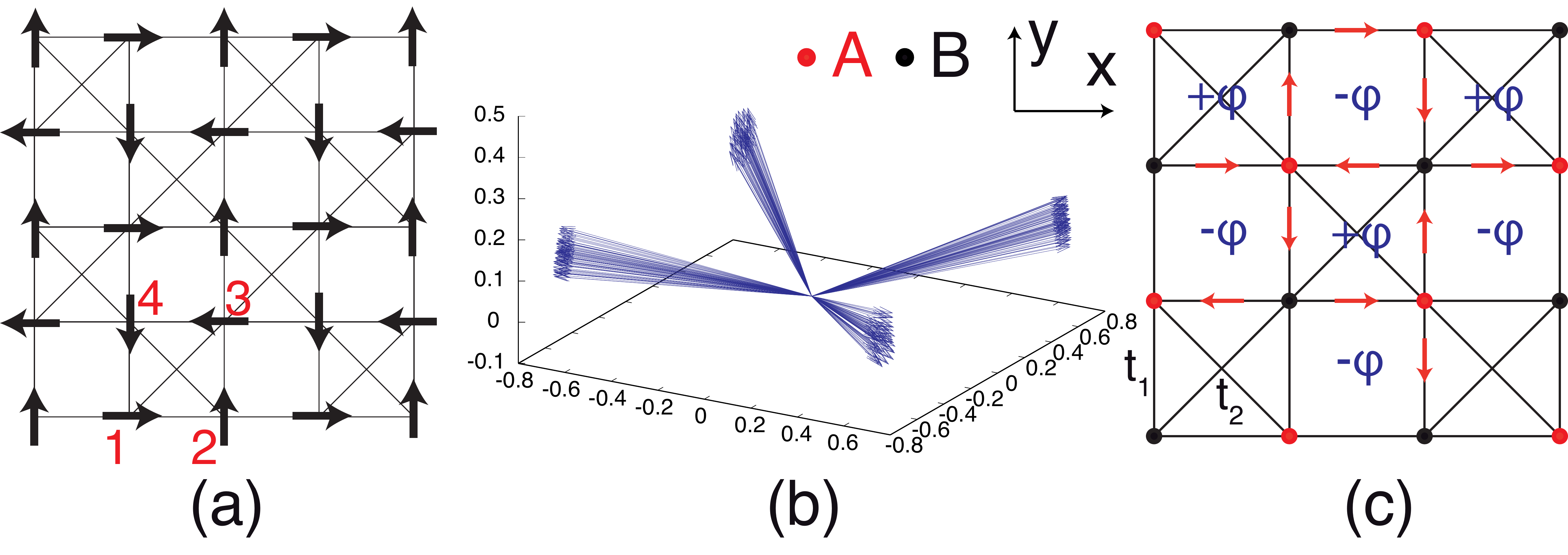}
\caption{\label{fig:fluxphase} (Color online) (a) Shows the localized spin texture of
  the ``flux'' phase; all spins are in plane, for number labeling see text. (b) Representation of the
  spins in the ``umbrella'' spin-chiral state, as obtained with
  MCMC+optimization for a 16x16 lattice with $J_{AF} = 0.105$, where
 $\delta = 0.148$ compared to $0.141$ as would be expected analytically; all spins have been translated to a single site.
  (c) Staggered flux arrangement on the checkerboard lattice
  corresponding to this insulating umbrella phase. Red arrows indicate
  the gauge choice of the ``gauge flux''. Note that the indicated flux
  threads the squares, so half of it threads the 
  triangles. } 
\end{figure}

{\it Model and method.---} 
We thus consider itinerant electrons coupled to localized core spins,
described by the canonical one-band double-exchange model with a
competing antiferromagnetic (AFM) super-exchange interaction, on the
checkerboard lattice at the density of one electron per two sites,
where the kinetic energy promoting ferromagnetic (FM) spin
correlations is strongest. The resulting Hamiltonian is  
\begin{equation} \label{eq:ham}
H = -\sum_{\langle ij\rangle} \, t_{ij} (\crferm_i\anferm_j + H.c.) 
+ J_{AF}\sum_{\langle ij\rangle} {\bf S}_i \cdot  {\bf S}_j ,
\end{equation}
where $\crferm_i$ ($\anferm_i$) creates (annihilates) a fermion on
site $i$. Here we have assumed for simplicity that an infinite Hund's
rule perfectly aligns the fermion spin to the localized spins  ${\bf
  S}_i$, but we have verified that the results presented in this
Letter remain valid also for large but finite coupling.  
As the onsite spins are classical and one can take without loss of generality $|{\bf S}_i| =1 $, they are completely specified by polar and azimuthal angles $(\theta_i,\phi_i)$. As hopping and superexchange are identical along the ``straight'' and ``diagonal'' edges of the checkerboard lattice, see Fig.~\ref{fig:fluxphase}(a), both types of bonds are included in the sum over $\langle ij\rangle$.    
The hopping amplitude depends on the core spins as $t_{ij} =
t[\cos(\theta_i/2)\cos(\theta_j/2)+\sin(\theta_i/2)\sin(\theta_j/2)e^{-i(\phi_i-\phi_j)}
]$~\cite{Dagotto:2001wb}. 
AFM super-exchange is given by $J_{AF}$ and all energies will be measured in units of 
the hopping amplitude $t$. We use Markov Chain Monte Carlo (MCMC)
simulations to treat the classical spins, where the weight of a spin
configuration is given by the free energy of the effective fermionic 
Hamiltonian, as obtained by exact diagonalization~\cite{Dagotto:2001wb}. We have performed
calculations on lattices with $N = 8^2$, $12^2$, $16^2$, and $20^2$ sites. MCMC
calculations were supplemented with an energy optimization in order to
suppress thermal fluctuations~\cite{Yu:Prb09}. 

\begin{figure}
\includegraphics[width=\columnwidth]{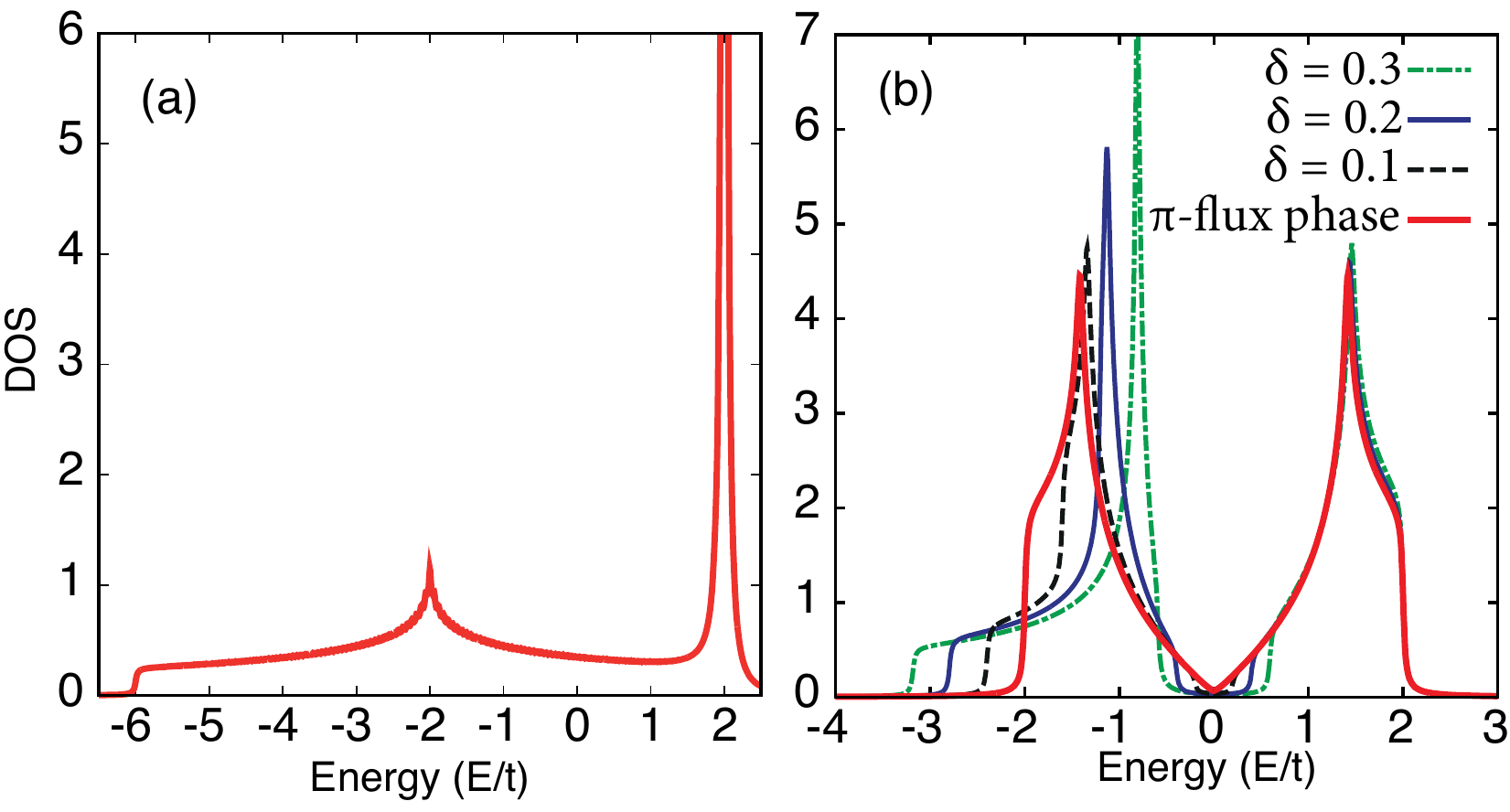}
\caption{\label{fig:DOS} (Color online) Electronic density of states of (a) the 
  FM (spinless) phase, (b) the ``flux'' phase (in red) and spin-chiral umbrella
  phase for various $\delta$ (see text).   
\label{fig:states}} 
\end{figure}

{\it Checkerboard lattice magnet.---} 
Let us
first discuss the two terms in the Hamiltonian separately,
corresponding to the limits $J_{AF} \rightarrow 0$ and $J_{AF}
\rightarrow \infty$, and turn to them competing at the next stage.
For vanishing super-exchange coupling $J_{AF} \to 0$, one finds a
saturated FM state, equivalent to free spinless
fermions on a checkerboard lattice. There are two bands
$E_+ = 2t$ and $E_- = -2t-4t\cos k_x\cos k_y$ and the 
density of states (DOS),  $D(\omega) 
= \langle \tfrac{1}{N} \sum_k \delta(\omega-\epsilon_k) \rangle$, is
shown in Fig.~\ref{fig:DOS}(a). The first Brillouin zone (BZ) is given by $(k_x,k_y) \in \{ |k_x+k_y|
\leq \pi\} \cap \{ |k_x-k_y| \leq \pi\}$. 
The Heisenberg term dominating for $J_{AF}
\rightarrow \infty$ is geometrically frustrated and supports an
infinitely large classical degeneracy of spin ground 
states. This can be seen by noting that it can be
rewritten in terms of the total spins on the crossed plaquettes ${\bf
  S}_{\mathcal{P}} = {\bf S}_1+{\bf S}_2+{\bf S}_3+{\bf S}_4$, giving 
\begin{equation}
J_{AF}\sum_{\langle ij\rangle} {\bf S}_i \cdot  {\bf S}_j =
\frac{J_{AF}}{2}\sum_{\mathcal{P}} {\bf S}_{\mathcal{P}} \cdot {\bf
  S}_{\mathcal{P}} - J_{AF}N  , 
\end{equation}
where $N$ is the number of sites. Clearly, the lowest energy is
obtained when ${\bf S}_{\mathcal{P}} = 0 $. This {\it local}
requirement on the spins of each crossed square is similar to the spin-ice rule in the
pyrochlore lattice. For Ising spins~\cite{Jaubert:2011fp} the
rule corresponds to ``two up-two down''. If noncollinear spin
arrangements are permitted, as we consider here, the class of states
${\bf   S}_{\mathcal{P}} = 0 $ is further increased.  

For dominant super-exchange coupling $J_{AF}\gg 1$, the magnetic order
is expected to belong to the highly degenerate ground-state manifold 
(GSM) fulfilling ${\bf S}_{\mathcal{P}} = 0 $. The kinetic energy can
leave the degeneracy intact, lift it partially or lift it completely,
singling out a unique non-degenerate spin arrangement. Our MCMC
calculations show that the latter is the case, the electrons pick out
a particular coplanar, but not collinear, state that is schematically depicted in
Fig.~\ref{fig:fluxphase}(a). Non-diagonal bonds connect orthogonal
spins, while diagonal bonds connect AFM spins, effectively
excluding them from the hopping term. Going around a square plaquette,
the electrons pick up a phase $e^{i\pi}$, corresponding to a
time-reversal invariant flux of $\pi$, and this special ``flux'' phase
has been shown to arise in models for high-Tc
superconductors~\cite{AFFLECK:1988vg,lorenzana} and in the double-exchange models on
the square
lattice~\cite{Yamanaka:1998uw,Agterberg:2000un,Aliaga:2001uy,Chen:2010kg}. On
the unfrustrated square lattice, it competes with the Neel state for strong
$J_{AF}$~\cite{Yamanaka:1998uw,Agterberg:2000un,Aliaga:2001uy}, but
since it fulfills the ice-rules, it remains stable for $J_{AF} > 0.12$
on the checkerboard lattice.

\begin{figure}
\includegraphics[width=\columnwidth]{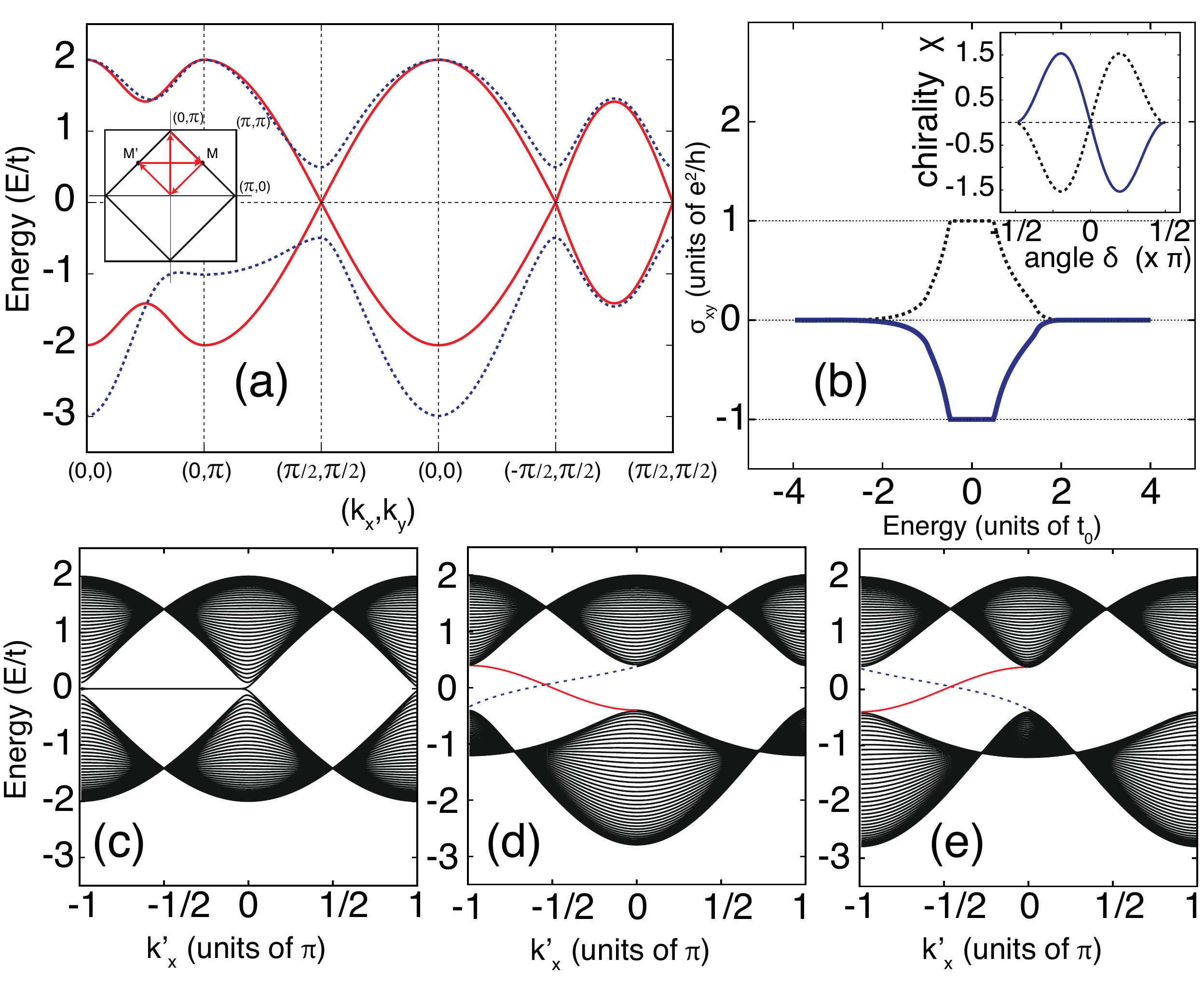}
\caption{(Color online) (a) The band structure of the gapless flux
  phase (red) and the insulating chiral phase (blue, $\delta=0.3$)
  along a path in the Brillouin zone specified in the inset. 
  (b) Quantized Hall conductivity in the chiral state ($\delta=0.3$), when
  the Fermi level is in the gap, the quantized value depends on the
  chirality of the spin state. The inset shows the calculated
  chirality of the states $\Lambda^{\pm}$, where dashed (solid)
  corresponds to $+$ ($-$). (c-e) Spectrum of the flux phase
  calculated for a strip geometry, which explicitly shows the edge
  states at the open boundary. (c) $\pi$-flux phase ($\delta=0.0$) exhibits
edge states similar to graphene. (d,e) Chiral gapped phase ($\delta=0.2$); chiral edge
  states connect valence and conduction bands. The states drawn with 
  solid (dashed) lines lives on the top (bottom) edge. The chirality
  in (e) is reversed with respect to (d), the right- and left-moving
  states are consequently exchanged. \label{fig:flux}} 
\end{figure}

The DOS of the flux phase shows semi-metallic behaviour [see
Fig.~\ref{fig:DOS}(b)] that originates from two Dirac
points in the spectrum. Low-energy excitations are described by a
relativistic Dirac equation, in full analogy
with graphene~\cite{CastroNeto:2009cl}. The core-spin
texture $\Lambda_i = (\theta_i,\phi_i)$ of the flux phase can be
written as $\Lambda_i = (\pi/2,(i-1)\pi/2 ) $ with $i=1,2,3,4$ [see
Fig.~\ref{fig:fluxphase}(a)]. Even
though the magnetic texture has a 4-site unit 
cell, the two-site electronic unit cell need not be enlarged. The electronic
Hamiltonian, in the $(\psi^\dagger_A, \psi^\dagger_B)$ basis, is then
given by $H({\bf k}) = {\bf d}({\bf k})\cdot
\boldsymbol{\sigma}$, with he Pauli matrices $\boldsymbol{\sigma} =
(\sigma^x,\sigma^y,\sigma^z)$ and ${\bf d}({\bf k}) = -(\cos k_x+\cos
k_y, \cos k_x-\cos k_y,0)$. The band structure of this state is shown
in Fig.~\ref{fig:flux}(a). The two inequivalent Dirac points, or
valleys, are located at ${\bf  M}_\pm = (\pm\pi/2,\pi/2)$. Expanding
around the Dirac points yields an effective low-energy Hamiltonian
with the two Dirac spinors   
%
$\Psi^\dagger_1  =  (\psi^\dagger_{A}({\bf  M}_++{\bf p}),\psi^\dagger_{B}({\bf  M}_++{\bf p}))$
and 
$\Psi^\dagger_2  =  (\psi^\dagger_{A}( {\bf M}_-+{\bf p}),\psi^\dagger_{B}( {\bf M}_-+{\bf p}))$.
Rotations on momentum and spinor
components bring us to 
the familiar form $H({\bf q}) = \nu^z \otimes (q_x\sigma^x +
q_y\sigma^y)$, where $\nu$-Pauli matrices act on the valley index. 
This is equivalent to graphene, with two valleys around 
which the electrons are described by the Dirac equation and an
interesting mapping exists between graphene and the flux 
phase~\cite{Hatsugai:2006db}.
One important difference to graphene is that we have here no spin
degeneracy, as the spin degree of freedom was integrated out by tying
the fermion spin to the core spins. 

{\it Massive QAH Dirac fermions.---} 
Having established that a not-too-large electronic kinetic energy
selects a unique non-collinear pattern for the checkerboard
double-exchange magnet, which has a graphene-like Dirac spectrum, we
consider next what happens upon an increase of the itineracy. 
%
%
Lowering $J_{AF}$, we find that the magnetic interactions enforcing the tetrahedron rules are
overcome by the electronic kinetic energy for $J_{AF}\lesssim 0.12$.
The transition is continuous and can be understood as a tilting of
the flux-pattern out of the plane, forming an "umbrella''. An
example is shown in Fig.~\ref{fig:fluxphase}(b): the spins 
fall along four directions, whose projections onto the $x$-$y$ plane
mirror the ``flux''-phase pattern, but there is an additional FM component
along the $z$ axis. The spins can be described using an Ising variable $s=\pm
1$ (which will turn out to correspond to a scalar
spin chirality) and a continuous parameter
$\delta$ giving the tilting along $\mp z$: $\{\Lambda^s_i(\delta) \}=
(\pi/2+\delta,s(i-1)\pi/2)$, where $i=1,2,3,4$ again runs around a crossed
plaquette. A similar scenario, but involving a more complex 8-site
unit cell and leading to Chern numbers $\pm 2$, arises on a square lattice with longer-range couplings when
nearest-neighbor hoppings are strongly modulated~\cite{Chen:2010kg}.

The scalar spin chirality of the state is defined as $\chi = \sum_{\mathcal{T}} {\bf
  S}_i \cdot {\bf S}_j \times {\bf S}_k $, where the sum is over all
triangles $\mathcal{T}$ of the checkerboard lattice, and ${\bf S}_i
\cdot {\bf S}_j \times {\bf S}_k$ is taken in the counter-clockwise
direction. The chirality as function of $\delta$ is plotted in the inset of
Fig.\ref{fig:states}(b) for umbrella states $\Lambda^\pm $, it is $\chi \approx
-s\delta$ for small $\delta$. The label $\pm s$ decides the sign of the
chirality for $\delta>0$ and is related to
a (counter-)clockwise rotation of the spin projection onto
the $x$-$y$ plane. The umbrella states, in addition to a continuous spin rotation symmetry,
thus also break a discrete $\mathbb{Z}_2$ symmetry. 
Since breaking a discrete
symmetry in $2D$ is possible at finite temperature, this opens up the
possibility for chiral ordering in absence of long-range magnetic
ordering~\cite{Kato:2010du}. 

The effect of the tilting on the electronic degrees of
freedom is to break time-reversal symmetry, as fluxes through elementary
plaquettes are related to the solid angle subtended by the spins
surrounding the plaquette. 
Calculating the hoppings in the umbrella states, we find that hopping
on the straight bonds is given by $ t_1^s = e^{-s
  i\pi/4}(1-s i\sin\delta)/\sqrt{2}  $, with $|t_1^s|=
\sqrt{(1+\sin^2\delta)/2 } \equiv t_1$ and $\phi^s_1  = \arctan
(-s\sin\delta) -s\pi/4\equiv \phi^s$ [see
Fig.~\ref{fig:fluxphase}(c)]. In addition, hopping along the diagonal
bonds is no longer 0 but $t_2 =- \sin \delta$, independent of
chirality. Using these expressions we write the 
effective Hamiltonian for the electrons as 
\begin{eqnarray} \label{eq:chiral}
H^s({\bf k}) &=& {\bf d}({\bf k})\cdot \boldsymbol{\sigma} + d^0({\bf k}) \sigma^0, \ \textrm{with} \\
d^0({\bf k}) &=& -2t_2 \cos k_x \cos k_y, \ d^3({\bf k}) = 2t_2\sin k_x \sin k_y\nonumber \\
d^1({\bf k}) &=& -2t_1\cos \phi^s(\cos k_x +\cos k_y ),\ \textrm{and}\nonumber \\
d^2({\bf k}) &=& -2t_1 \sin \phi^s(-\cos k_x +\cos k_y ),\nonumber
\end{eqnarray}
where the two states referred to by the matrices are again the two sites of the unit cell and $\sigma^0$ is the unit matrix so that the Dirac Hamiltonian above is recovered for $\delta=0$, implying $\phi^s=\pi/4$ and $t_2=0$.
From the DOS [Fig.~\ref{fig:DOS}(b)] and the
band structure [Fig.~\ref{fig:flux}(a)], it is clear that finite $\delta\neq 0$ opens a
gap for the Dirac cones. 
Since the hoppings are complex and the diagonal bonds have been activated, 
both time-reversal and parity symmetries are broken, allowing a QAH
state~\cite{Onoda:2002vo}. To establish that the gapped state is 
indeed topologically non-trivial, we calculate the Chern number $C_n =
\frac{1}{2\pi i} \oint_{\partial BZ} d{\bf k}\cdot {\bf A}({\bf k})$,
where ${\bf A}({\bf k}) = \langle n{\bf k}|\nabla_{{\bf k}} |n{\bf k}
\rangle$ is the Berry connection and find $C =
\text{sgn}(t_2)\text{sgn}(\sin 2\phi^s)$. Chirality and Chern number
hence perfectly correlate and we observe that inverting the magnetic
polarization $\delta \rightarrow -\delta$ flips both the spin chirality
and the Chern number. The off-diagonal Hall conductivity 
as a function of chemical potential, obtained from
Eq.~(\ref{eq:chiral}) for $\delta = 0.3$, is shown in
Fig.~\ref{fig:flux}(b). Figure~\ref{fig:flux}(c-e) shows the effect of 
non-trivial topology on the edge of the system: chiral edge states
connect valence and conduction band. As can be seen by comparing
Figs.~\ref{fig:flux}(d) and~\ref{fig:flux}(e), the direction of the
edge currents can be reversed by inverting the spin chirality. The
latter can be easily manipulated by a small magnetic field.

The observation that spin configurations of the umbrella states are continuously
connected to the coplanar flux phase suggests that the
electronic QAH state can be understood from the low-energy physics at
the Dirac points. 
We will demonstrate this by analyzing, in the spirit of Ref.~\cite{HALDANE:1988uf}, the system in a presence of an external magnetic field $B$ and then take the limit of $B\rightarrow 0$. 
Focusing on the low-energy
theory, $\delta {\bf k} = {\bf k} - {\bf M}_\gamma$ ($\gamma = \pm$),
we introduce the magnetic field by way of a Peierls substitution $\hbar
\delta {\bf k} \rightarrow \hat{{\boldsymbol{ \pi} }}$, where
$\hat{{\boldsymbol{ \pi} }}$ is the dynamical momentum whose
components satisfy the commutation relation $[\momx,\momy] = i eB
\hbar$. We obtain two independent Hamiltonians for the two Dirac
points, $H_{\gamma} = v_F (\hat{\pi}^1_{\gamma}\sigma^x +
\hat{\pi}^2_{\gamma}\sigma^y ) + m_{\gamma}\sigma^z$, which indeed has
the appearance of the relativistic Dirac equation in a magnetic
field. As can be seen by comparing to Eq.~(\ref{eq:chiral}), our mass
term $m_\gamma = 2 \gamma t_2$ is a direct consequence of finite
$t_2=-\sin \delta$, and thus of finite chirality $\delta\neq
0$. Operators $\hat{\pi}^1_{\gamma}$ and $\hat{\pi}^2_{\gamma}$ are 
derived from $\hat{{\boldsymbol{ \pi} }}$ and satisfy the commutation
relation $[\hat{\pi}^1_{\gamma},\hat{\pi}^2_{\gamma}] = -i\gamma
\sin(2\phi^s) eB \hbar$. Relativistic Dirac fermions in a magnetic field
are known to exhibit zero modes in their
spectrum~\cite{JACKIW:1984wi}, which cause the charge density
imbalance in the ground state, potentially leading to an integer
QAH effect. Here, the zero modes have energy $E_{0,\gamma}
= -\gamma m_\gamma \text{sgn}(\sin 2\phi^s) \text{sgn}(eB)$, and the spectrum is
asymmetric when $m_+$ and $m_-$ have opposite sign. Following Haldane,
we obtain the
off-diagonal conductivity in the limit $B\rightarrow 0$,  $\sigma_{xy}
= \nu e^2/h$, where $\nu = \frac{1}{2}\text{sgn}(\sin 2\phi^s)
[\text{sgn}(m_+)-\text{sgn}(m_-)] = \text{sgn}(t_2)\text{sgn}(\sin
2\phi^s)$. Hence, the gapped QAH umbrella state can be interpreted as
Dirac fermions becoming massive, with masses of \emph{opposite}
sign, indeed, the $d^z(\bf k)$ component of Eq.~(\ref{eq:chiral}) has
opposite sign at the two Dirac points $(\pm \pi/2,\pi/2)$. A
sublattice potential, which also gaps out the Dirac fermions, 
would in contrast lead to equal masses, and the edge states would not
cross the chemical potential. 

{\it Discussion and conclusions.---} 
We investigated the interplay of itinerant
electrons with a frustrated AFM spin background on the
checkerboard lattice using Monte-Carlo methods.
 From the macroscopically degenerate {\it spin-ice}
ground-state manifold, which optimizes the AFM interactions, the electron kinetic
energy selects a unique magnetic ground state. The low-energy
electronic states of the selected $\pi$-flux phase are given by a
relativistic Dirac equation, and slightly stronger kinetic energy
induces a spin chirality, from which the Dirac fermions inherit a
topologically nontrivial mass. The Kondo-lattice model on the
checkerboard model thus provides a direct realization of Haldane's
proposal for obtaining a QAH state~\cite{HALDANE:1988uf}. In addition,
the QAH state's chirality is coupled to a FM spin polarization and the
direction of the edge currents can thus be switched by a 
magnetic field, an alluring property for quantum spintronics applications. 

Our findings are also relevant in the broader context of
fractionalization of quantum numbers in two different ways. First, the
QAH state on the checkerboard lattice is actually a prominent 
candidates for hosting a {\it fractional} quantum-Hall--like state without a
magnetic field, because the topologically nontrivial band can be made
almost flat by tuning hoppings and flux~\cite{Sun:2011dk,Sheng:2011iv,Regnault11}. It turns out that even if the
flux and effective hoppings 
$t_1$ and $t_2$ that emerge in the ``umbrella'' configuration do not
lead to very flat bands, additional longer-range hopping $-2t _3(\cos
2k_x+\cos 2k_y)$ can give a ratio of band gap vs. band width of
$\approx 5$ for $\delta=0.3$, considerably less than ratios achievable by tuning
all parameters~\cite{Sun:2011dk} or in $t_{2g}$-orbital
systems~\cite{Venderbos:11_flat,2011arXiv1109.5955V}, but comparable
to $e_g$~\cite{Venderbos:11_flat} systems or a square-lattice
model~\cite{neupert10}. Second, it was recently demonstrated that
vortex defects of the localized magnetic order underlying a QAH state can carry
fractional charge and spin quantum numbers in the electronic
sector~\cite{Muniz:2011ti}.  

{\it Acknowledgements.---} 
This research was supported by the Interphase Program of the Dutch Science Foundation NWO/FOM (JV and JvdB) and by the Emmy-Noether program of the DFG (MD). SK acknowledges support from DST, India.

\bibliographystyle{prsty-etal}
\bibliography{arXiv_submission.bbl}

\end{document}